\documentclass[12pt,a4paper]{article}
\usepackage{amssymb,amsmath}
\newcommand{\Slash}[1]{\ooalign{\hfil/\hfil\crcr$#1$}}
\newfont{\bg}{cmr10 scaled\magstep3}
\newcommand{\Bzero}{\smash{\raise1ex\hbox{\bg 0}}}
\newcommand{\ncpl}{{\mathbb T}^2_\frac{2}{N}}
\newcommand{\Integer}{\mathbb Z}
\newcommand{\Tr}{\mbox{\rm Tr}}

\begin{document}

\title{Large $N$ Structure of the IIB Matrix Model}
\author{{\sc Naofumi Kitsunezaki}\thanks{Email address:
kitsune@eken.phys.nagoya-u.ac.jp} ~and
{\sc Shozo Uehara}\thanks{Email address:
uehara@eken.phys.nagoya-u.ac.jp}\vspace{4mm}\\
{\it Department of Physics, Nagoya University,}\\
{\it Chikusa-ku, Nagoya 464-8602, Japan}}
\date{}
\maketitle
\vspace{-70mm}
\begin{flushright}
	DPNU-00-36\\
	hep-th/0010038\\
	October 2000
\end{flushright}
\vspace{50mm}

\begin{abstract}
We study large $N$ behavior of the IIB matrix model using the
equivalence between the IIB matrix model for finite $N$ and a field
theory on a non-commutative periodic lattice with $N\times N$ sites.
We find that the large $N$ dependences of correlation functions can be
obtained by naively counting the number of fields in the field
theory on a non-commutative periodic lattice.
Furthermore, the large $N$ scaling behavior of the coupling constant
$g$ is determined if we require that the expectation values of
Wilson loops be calculable.
\end{abstract}

\section{Introduction}
Four years ago, large $N$ matrix models were proposed as
superstring theories \cite{BFSS,IKKT,FKKT,AIKKT}.
The IIB matrix model \cite{IKKT} is the zero volume limit \cite{EK} of
a ten-dimensional super Yang-Mills theory.
The gauge fields in the IIB matrix model are written as $N\times N$
hermitian matrices.
The eigenvalues of these matrices are interpreted as space-time
coordinates. This model represents an open space-time, because the
eigenvalues of the matrices can take values in the infinite region.
Some studies of a closed space-time with periodic boundary conditions
have been done by changing the hermitian matrices to unitary matrices
\cite{KN}. When $N=2$, we can successfully integrate the fermions by
hand and analytically investigate the model with the bosonic degrees
of freedom \cite{TsuSu}. However, it is almost impossible to do
something like this when $N>2$.

In the IIB matrix model, space-time coordinates are written in terms
of the gauge fields and they do not commute with each other, in
general.
In ref.\cite{CDS} toroidal compactifications of the matrix models on
a non-commutative torus were investigated. Since then,
non-commutative geometries in string and matrix theories have been
vigorously studied.

The commutation relations of the fields in non-commutative space-time
are very similar to the commutation relations of $U(N)$ algebra, which
have been investigated in the context of supermembranes.
\cite{deWit,poppe,zachos,castro}.
In fact, it is known that there is a one-to-one correspondence between
the action of an $N\times N$ matrix model and that of the field theory
on a non-commutative periodic lattice (NCPL) with $N\times N$ sites,
which we write as $\ncpl$ \cite{bars,tokyou,landi}.
Since the commutation relations of coordinates of the NCPL are
proportional to $\frac{1}{N}$, we expect that the matrix model in
the large $N$ limit has a one-to-one correspondence to a field theory
on the continuous torus.

The naive large $N$ limit of the IIB matrix model is the Schild action
on a torus \cite{Schild}. For this reason, the Schild action on a torus
is a plausible candidate for the large $N$ limit of the IIB matrix
model.
However, we find that the large $N$ dependences of some expectation
values in the IIB matrix model for finite $N$ are different from those
in the regularized Schild theory on a periodic $N\times N$ lattice.
Since systems with infinite numbers of degrees of freedom are usually
described by field theories, and the large $N$ limit of the IIB matrix
model is a system with an infinite numbers of degrees of freedom, it
is natural to search for a field theory of the IIB matrix model in the
large $N$ limit.
Based on the guiding principle that the $N$ dependence for some
expectation values is the same as in the IIB matrix model, we consider
field theories on a torus.

We investigate the large $N$ dependences of the correlation functions
of the IIB matrix model using the field theory on an NCPL.
We find that the $N$ dependences of some correlation functions
calculated in the matrix model can be straightforwardly given with the
field theory on the NCPL using kinematical arguments.
We also show that correlation functions such as
\begin{equation}
	\left\langle\frac{1}{N}\Tr F(A,\Psi)\right\rangle
\end{equation}
can have non-trivial values in the large $N$ limit if we demand
$g^2N=O(N^0)$.

This paper is organized as follows. In section \ref{sec:MtoF}
we review the isomorphic mapping from a $U(N)$ matrix model to a field
theory on the NCPL.
In section \ref{sec:FNIIB} we discuss the large $N$ behavior of
correlation functions using the field theory on the NCPL, which is
isomorphic to the IIB matrix model.
In section \ref{sec:LNIIB} we discuss the field theory which
corresponds to the IIB matrix model in the large $N$ limit.
We show that the IIB matrix model cannot lead to the Schild action
\cite{Schild} straightforwardly.
The final section is devoted to conclusion and discussion.

\section{Mapping between the $U(N)$ Algebra and a Field on\\
a Non-Commutative Periodic Lattice}\label{sec:MtoF}
There are some investigations of the large $N$ limit of the $SU(N)$
algebra \cite{deWit,poppe,zachos}.
The $U(N)$ algebra\footnote{We assume $N$ is odd for definiteness.} is
generated by the following two matrices \cite{zachos}:
\begin{equation}
U=\left(
    \begin{array}{@{\,}ccccc@{\,}}
	1&&&&\\
	&e^{i\frac{4\pi}{N}}&\multicolumn{3}{c}\Bzero\\
	&&\ddots&&\\
	\multicolumn{3}{c}\Bzero &&e^{{i\frac{4\pi}{N}}(N-1)}\\
   \end{array}\right),~~~
V=\left(
    \begin{array}{@{\,}ccccc@{\,}}
	0&1&&&\\
	&0&1&\multicolumn{2}{c}\Bzero\\
	&&\ddots&\ddots&\\
	&\Bzero&&0&1\\
	1&&&&0
    \end{array}\right).
\label{eqn:gANDh}
\end{equation}
These matrices satisfy the relations
\begin{eqnarray}
	U^N&=&V^N~=~{\bf 1}\,, \label{eqn:rootN}\\
	V\,U&=&e^{i{\frac{4\pi}{N}}}U\,V\,.
\end{eqnarray}
The bases of the $U(N)$ algebra can be written
\begin{equation}
    {\bf T}_{(m_1,m_2)}=e^{i{\frac{2\pi}{N}}m_1m_2}\,U^{m_1}\,V^{m_2}.
\end{equation}
The commutation relations of ${\bf T}_{\bf m}
(\equiv {\bf T}_{(m_1,m_2)})$ are given by \cite{moyal,weyl}
\begin{equation}
	[ {\bf T}_{\bf m},{\bf T}_{\bf n} ]
	=-2i\sin\left({\frac{2\pi}{N}}\epsilon^{ab}m_an_b\right)
	{\bf T}_{\bf m+n}. \hspace{3mm}
	(\epsilon^{12}=-\epsilon^{21}=1,~a,b=1,2) \label{eqn:sinalg}
\end{equation}
This type of commutation relation is satisfied by plane-wave
functions on the two-dimensional non-commutative torus on which the
coordinates satisfy \cite{CDS}
\begin{equation}
	[x_1,x_2]=i{\frac{4\pi}{N}}. \label{eqn:2Dcr}
\end{equation}

We define the operator
$\Delta(\sigma)~(\sigma=(k^1/N,k^2/N),\, k^a\in \{0,1,\cdots,N-1\} =
\Integer_N)$, by
\begin{equation}
    \Delta(\sigma)={\frac{1}{N}}\sum_{m_a\in\Integer_N}\,
    {\bf T}_{\bf m}\,e^{-2\pi i\,{\bf m\cdot\sigma}},\label{eqn:invft}
\end{equation}
which maps $N\times N$ hermitian matrices to functions on the
two-dimensional $N\times N$ lattice with periodic boundary
conditions. The discrete space where $\sigma$ takes values is written
$\ncpl$.
Note that $\Delta(\sigma)$ satisfies
\begin{equation}
	\Delta((\sigma^1 +
	\ell^1,\sigma^2+\ell^2))=\Delta((\sigma^1,\sigma^2))\,.
	\hspace{10ex} (\ell^a\in\Integer)
\end{equation}
Equation(\ref{eqn:invft}) represents an inverse Fourier transformation
of the matrices ${\bf T}_{\bf m}$ which have momenta $2\pi{\bf m}$.
Using this $\Delta(\sigma)$, a hermitian matrix $A$ is mapped to a
field on $\ncpl$ according to
\begin{equation}
	A(\sigma)=\Tr(\Delta(\sigma)A). \label{eqn:MtoF}
\end{equation}
Also, the product of two matrices $A$ and $B$ is mapped to the diamond
product \cite{bars} of two fields on $\ncpl$ according to
\begin{equation}
    \Tr(\Delta(\sigma)AB) \equiv A(\sigma)\diamond B(\sigma)
	= \frac{1}{N^2}\,\sum_{\tau,\omega\in\ncpl}\,
	e^{2\pi i N\,\epsilon^{ab} (\sigma - \tau)_a\,
	(\sigma - \omega)_b}\, A(\tau)\,B(\omega)\,.
	\label{eqn:TwoMTRX}
\end{equation}
Equation(\ref{eqn:MtoF}) can be rewritten as
\begin{equation}
    A(\sigma)= \sum_{\bf m}\{\frac{1}{N}\,\Tr\,({\bf T_{N-m}}A)\}\,
	e^{-2\pi i ({\bf N-m})\cdot \sigma} \equiv
      \sum_{\bf m} A_{\bf m}\,e^{2\pi i {\bf m}\cdot \sigma}\,,
\end{equation}
and using these Fourier transformed modes, $A(\sigma)\diamond
B(\sigma)$ in eq.(\ref{eqn:TwoMTRX}) is given by,
\begin{eqnarray}
    A(\sigma)\diamond B(\sigma)&=&
      \sum_{\bf m,n}\,e^{-\frac{2\pi i}{N}\,\epsilon^{ab} m_an_b}\,
      \left(A_{\bf m}\,e^{2\pi i{\bf m}\cdot\sigma}\right)\,
      \left(B_{\bf n}\,e^{2\pi i{\bf n}\cdot\sigma}\right)\nonumber\\
    &\equiv&A(\sigma)~e^{ \frac{i}{2\pi N}\,\epsilon^{ab}\,
	{\overleftarrow\partial}_a{\overrightarrow\partial}_b}~B(\sigma)\,,
	\label{eqn:TwoMXFT}
\end{eqnarray}
where the partial derivative $\partial_a$ is used symbolically, but
it becomes a normal derivative in the large $N$ limit.
Then the diamond product is a discrete-space version of the star
product \cite{bars}.

Let us consider the action of matrices $A^i$,
\begin{equation}
	S= \Tr F(A^i). \label{eqn:Smat}
\end{equation}
The $\Delta$ mapping (\ref{eqn:MtoF}) maps this action (\ref{eqn:Smat})
to
\begin{equation}
	S={\frac{1}{N}}\sum_{\sigma\in\ncpl}F_\diamond(A^i(\sigma)),
\label{eqn:Sgen}
\end{equation}
where $F_\diamond(\cdot)$ denotes the quantity whose functional form is
the same as that of $F(\cdot)$ but in which all the products of
matrices are replaced by the diamond products of the corresponding
fields on $\ncpl$.
The measure of the path integral is given by
\begin{equation}
	{\cal D}A^i(\sigma)=\prod_{0\leq m_a < N}dA^i_{\bf m}.
\end{equation}

To this point we have considered a $U(N)$ matrix model, but it is
obvious that we can follow the same procedures for a $SU(N)$ matrix
model. Since matrices are traceless, the resultant fields on $\ncpl$
have no zero mode.

\section{Field Theory on a Non-commutative Periodic\\
Lattice as a IIB Matrix Model}\label{sec:FNIIB}
We now apply the mapping in the previous subsection to the IIB matrix
model \cite{IKKT}. The action of the IIB matrix model \cite{IKKT}
is given by
\begin{equation}
    S^M_{\mbox{\scriptsize IIB}}=
	-{\frac{1}{g^2}}\,\Tr\left({\frac{1}{4}}[A^\mu,A^\nu]^2
	+{\frac{1}{2}}\,{\overline\Psi}\,\Gamma_\mu[A^\mu,\Psi]\right),
\label{eqn:SikktOrg}
\end{equation}
where $A_{\mu}~(\mu = 1,2,\dots,10)$ and $\Psi$ are $N\times N$
hermitian matrices and $\Psi$ are ten-dimensional Majorana-Weyl
spinors.
Then the corresponding action of the field theory on $\ncpl$ is
given by (cf. eqs.(\ref{eqn:Smat}) and (\ref{eqn:Sgen}))
\cite{tokyou,landi}
\begin{equation}
  S_{\mbox{\scriptsize IIB}_N}=-{\frac{1}{g^2N}}\sum_{\sigma\in\ncpl}
	\left({\frac{1}{4}}\,[A^\mu(\sigma),A^\nu(\sigma)]_\diamond^2
	+{\frac{1}{2}}\,{\overline\Psi}(\sigma)\diamond\Gamma_\mu
	[A^\mu(\sigma),\Psi(\sigma)]_\diamond\right)\label{eqn:SikktNCT}.
\end{equation}

We now consider the $N$ dependence of correlation functions with the
field theory on the NCPL (\ref{eqn:SikktNCT}) and we show that
the results obtained from Monte Carlo simulations
\cite{MCsimu,AABHN,alpha} and perturbative investigations
\cite{MCsimu} are straightforwardly derived.
The starting point is the fact that the expectation values of the
bosonic and the fermionic parts of the action are proportional to
$N^2$, respectively \cite{MCsimu}.
We explain this point using the bosonic part of the action here.
Let us first define the partition function with a parameter $\kappa$,
\begin{eqnarray}
  Z[\kappa]&=&\int {\cal D}A{\cal D}\Psi\,
	e^{\frac{1}{g^2N}\sum\limits_\sigma
	\left(\frac{\kappa}{4}\,
	[A^\mu(\sigma),A^\nu(\sigma)]^2_\diamond
	+ \frac{1}{2}{\overline\Psi(\sigma)}\,\diamond\,\Gamma_\mu
	[A^\mu(\sigma),\Psi(\sigma)]_\diamond \right)}\nonumber\\
   &=&\kappa^{-(\frac{D}{4}+ 2^{[\frac{D}{2}]-4}) (N^2-1)}
	\,Z[\kappa=1].\label{eqn:Z-kappa}
\end{eqnarray}
Next, we differentiate $Z[\kappa]$ with respect to $\kappa$ and take
the $\kappa\rightarrow1$ limit. Then the expectation value of the
bosonic part of the action is given by
\begin{equation}
    \left\langle -\frac{1}{4g^2N}
	\sum_{\sigma\in \ncpl}
	[A^\mu(\sigma),A^\nu(\sigma)]^2_\diamond\right\rangle
    = \left(\frac{D}{4}+2^{[\frac{D}{2}]-4}\right)\,(N^2-1)\,.
	\label{eqn:com-dist}
\end{equation}
If we assume that the order of $N$ is not affected by
the commutator and that the summation over $\sigma$ gives
$\sum\limits_{\sigma} = O(N^2)$ \cite{KU}, we have
\begin{equation}
    A^\mu(\sigma) = O((g^2N)^{\frac{1}{4}}).\label{eqn:scaling-b}
\end{equation}
We can make a similar argument to the fermionic part of the action,
and we obtain
\begin{equation}
  \Psi(\sigma) = O((g^2N)^{\frac{3}{8}})\,.\label{eqn:scaling-f}
\end{equation}
{}From eq.(\ref{eqn:scaling-b}), we can estimate the $N$ dependence of
correlation functions as
\begin{equation}
  \left\langle {\frac{1}{N}}\Tr (A^{\mu_1}\cdots A^{\mu_{2k}})
	\right\rangle
  =\left\langle{\frac{1}{N^2}}
	\sum_{\sigma\in \ncpl} \Big(A^{\mu_1}(\sigma)
	\diamond\cdots\diamond A^{\mu_{2k}}(\sigma)\Big)\right\rangle
	=  O((g^2 N)^{\frac{k}{2}})\,.\label{eqn:predictA}
\end{equation}
As a special case of (\ref{eqn:predictA}), the distribution of
the eigenvalues of the bosonic matrices is estimated as
\begin{equation}
  \left\langle\frac{1}{N}\,\Tr (A^2)\right\rangle
	= O(gN^{\frac{1}{2}})\,.\label{eqn:trAA}
\end{equation}
Furthermore, we can also estimate the $N$ dependences for any
correlation functions including fermions.
Let $F_{(m,n)}[A,\Psi]$ be a homogeneous polynomial
of terms of order $m$  in $A^\mu$ and order $n$ in $\Psi$.
For example, $F_{(1,2)}[A^\mu,\Psi]={\overline\Psi}{\Slash A}\Psi$.
Then we obtain
\begin{equation}
	\left\langle\frac{1}{N}
	\Tr \left(F_{(m,n)}[A^\mu,\Psi]\right)\right\rangle
	=O\bigl((g^2N)^{\frac{1}{4}m+\frac{8}{3}n}\bigr).\label{eqn:TrF}
\end{equation}
We consider Wilson loops,
\begin{equation}
  \left\langle\left[{\frac{1}{N}}\,\Tr \left( P\,
	e^{i\int d\sigma(k(\sigma)\cdot A+{\overline\lambda}(\sigma)\Psi)}
	\right)\right]\cdots\right\rangle\,,\label{eqn:WilsonLoop}
\end{equation}
which are not homogeneous polynomials. However, we will see that
they are finite if $g^2N$ is fixed in the large $N$ limit.
{}From refs. \cite{MCsimu,alpha} and \cite{AABHN} it is known
that any Wilson loops are finite if $g^2N$ is fixed in the large $N$
limit.\footnote{The correlation functions of the Wilson loops are not
the connected parts in our case. It is shown in ref.\cite{AABHN} that
connected $n$ point Wilson loops are $O(N^{-(2n-1)})$ for the bosonic
model and $O(N^{-n})$ for the supersymmetric model. These orders are
due to the factorization property and our result does not contradict
these results.}
We have studied this from the kinematical point of view, and
it is interesting that our naive argument reproduce the
results obtained using Monte Carlo simulations.
This suggests that investigating the IIB matrix model from the NCPL
point of view will be useful.

It is obvious that we can follow the same arguments with
the bosonic model, for which
\begin{equation}
  S_b=-\frac{1}{4g^2}~\Tr ([A^\mu,A^\nu]^2)\,,\label{eqn:Bmodel}
\end{equation}
whose large $N$ limit is shown to exist in ref.\cite{KS}, and it gives
the same results for the $N$ dependences as the supersymmetric model.
(\ref{eqn:SikktOrg})  Hence, for example, eq. (\ref{eqn:trAA}) holds.
In refs. \cite{MCsimu} and \cite{alpha}, some correlation functions
were calculated with Monte Carlo simulations and the results there agree
with ours.

\section{Large $N$ Limit of the IIB Matrix Model}\label{sec:LNIIB}
It would be quite difficult to write down the action of the IIB matrix
model in the large $N$ limit directly from the original action
(\ref{eqn:SikktOrg}).
However, eq. (\ref{eqn:SikktNCT}), the field theory action on the
NCPL, resembles the Schild action \cite{Schild}, and hence one may
think that the Schild action would be the correct action in this
limit.
In fact the following arguments support such expectations.
First, eq.(\ref{eqn:2Dcr}), the commutator between two coordinates of
the NCPL,  vanishes in the $N\rightarrow0$ limit.
This means that a field theory on the NCPL (\ref{eqn:SikktNCT})
becomes a field theory on an ordinary commutative continuous torus.
Next, the commutation relations of fields with respect to the diamond
product are given by
\begin{eqnarray}
  [A(\sigma),B(\sigma)]_\diamond&=&
	2i\,A(\sigma)\,\sin\left({\frac{1}{2\pi N}}\,\epsilon^{ab}
	{\overleftarrow\partial_a}{\overrightarrow\partial_b}\right)
	B(\sigma)\nonumber\\
   &=&{\frac{i}{\pi N}}\,\epsilon^{ab}\,\partial_a A(\sigma)\,
	\partial_b B(\sigma) +
	O\left(\frac{1}{N^2}\right)\,,\label{eqn:lNcr}
\end{eqnarray}
where the partial derivative in the first term on the r.h.s. may be
regarded as the difference on a periodic lattice.
In the large $N$ limit we could neglect $O(\frac{1}{N^2})$ terms in
(\ref{eqn:lNcr}), and then the action of the IIB matrix model
(\ref{eqn:SikktNCT}) would become
\begin{eqnarray}
    S_{\mbox{\scriptsize IIB}}&=&
    \frac{1}{g^2N^3}\,\sum_{\sigma\in\Integer^2_N}\,
    \left[\frac{1}{4}\,(\epsilon^{ab}\,\partial_aa^\mu (\sigma)\,
    \partial_ba^\nu(\sigma))^2 -\frac{i}{2}\,{\overline\psi}(\sigma)\,
    \epsilon^{ab}\,\partial_a{\Slash a}(\sigma)\,
    \partial_b\psi(\sigma)\right]\,,\label{eqn:Sikkt-sum}\\
  &=&\frac{1}{g^2 N} \int_{T^2}d^2\sigma \left[\frac{1}{4}\,
    (\epsilon^{ab}\,\partial_a a^\mu(\sigma)\,\partial_b
    a^\nu(\sigma))^2 -{\frac{i}{2}}\,{\overline\psi}(\sigma)\,
    \epsilon^{ab}\,\partial_a {\Slash a}(\sigma)\,
    \partial_b\psi(\sigma)\right]\,,\label{eqn:Sikkt}
\end{eqnarray}
where
\begin{eqnarray}
    a^\mu(\sigma)&=&\pi^{-\frac{1}{2}} A^\mu(\sigma)=
	\pi^{-\frac{1}{2}}\,\Tr(\Delta(\sigma)A^\mu)\,,\nonumber\\
    \psi(\sigma)&=&\pi^{-\frac{1}{4}}\Psi(\sigma)
	=\pi^{-\frac{1}{4}}\,\Tr(\Delta(\sigma)\Psi)\,.\label{eqn:IIBMtoF}
\end{eqnarray}
This means that the IIB matrix model of finite $N$ could be
approximated by the lattice-regularized Schild action.
The size of the matrix, $N$, is the same number as the number of sites
for each direction.
This argument is the inverse of that for the matrix regularization of
the Schild action \cite{Schild} using the NCPL.

Despite the above arguments, we show that the Schild theory
(\ref{eqn:Sikkt}) is not the large $N$ limit of the IIB matrix model.
First, we have seen that the large $N$ behavior of the fields
(\ref{eqn:scaling-b}) and (\ref{eqn:scaling-f}) depends on the $N$
dependence of the action.
However, the $N$ dependence of the IIB matrix model (\ref{eqn:SikktNCT})
is different from that of the lattice-regularized Schild
action (\ref{eqn:Sikkt-sum}). In fact, following arguments similar to
there given in the previous section, we find that the action
(\ref{eqn:Sikkt-sum}) gives
\begin{equation}
  \left\langle\frac{1}{N}\,\Tr (A^2)\right
	\rangle_{\mbox{\scriptsize Schild}} = O(gN^{-\frac{1}{2}})\,,
\end{equation}
Therefore we must conclude that (\ref{eqn:Sikkt}) itself, obtained in
the naive large $N$ limit, is not the large $N$ limit of the IIB
matrix model.

What is wrong with the argument that the Schild action
(\ref{eqn:Sikkt}) can be regarded as the large $N$ limit of the IIB
matrix model? The crucial oversight of such arguments is that we have
neglected $O(\frac{1}{N^2})$ terms in (\ref{eqn:lNcr}).
This is justified only if the Kaluza-Klein momenta on the
torus, ${\bf m}$, is sufficiently small compared to $N$.
The assumption in the previous section, that the order of $N$ is not
affected by the commutation relations when estimating the large
$N$ dependence \cite{KU}, implies that the configurations of large
momenta of order $N$ play a crucial role in calculating correlation
functions.

\section{Conclusion and Discussion}\label{sec:con}
We have taken notice of the equivalence of the IIB matrix model
(\ref{eqn:SikktOrg}) and the field theory on an NCPL
(\ref{eqn:SikktNCT}).
Specifically, a matrix $A$ is given by $\frac{1}{N} \sum\limits_{\sigma}
A(\sigma) \Delta(\sigma)$ in the coordinate representation and
$\sum\limits_{\bf m} A_{\bf m} {\bf T_m}$ in the momentum
representation, and eq. (\ref{eqn:SikktNCT}) is the action
in the coordinate representation.
We found that we can easily determine the large $N$ dependences
for correlation functions using the action in the coordinate
representation. We have shown that any correlation function
$\langle\frac{1}{N}\Tr F(A,\Psi)\rangle$ can have finite and
nontrivial values if we take $g^2N \sim O(N^0)$.
It is interesting that the connected parts of the
Wilson loops are renormalizable in the supersymmetric model when the
same limit is taken \cite{AABHN}.
We need an effective theory of the Wilson loops to obtain
the string field theory from the IIB matrix model \cite{FKKT}.
It is natural to demand that no correlation functions like
(\ref{eqn:TrF}) diverge from the field theoretical point of view.
This leads to the following double scaling limit:
\begin{eqnarray}
	N &\rightarrow& \infty,\nonumber\\
	g &\rightarrow& 0.\label{eqn:g2nfix}\hspace{10ex}
	(g^2 N : {\rm fixed})
\end{eqnarray}
We should note, however, that Eq. (\ref{eqn:g2nfix}) raises a question
about scales. If we fix $g^2N$, then ${\alpha'}^2\sim g^2N$ \cite{FKKT}.
On the other hand, $\left\langle\frac{1}{N}\Tr A^2\right\rangle$, which
is often interpreted as the extent of spacetime, also behaves like
$g^2N$. Then we must know the origin of this huge hierarchical
difference between scales of string and spacetime.
We may take another limit. Since  $\left\langle\frac{1}{N}\Tr
A^2\right\rangle\sim g^2N$ is the extent of spacetime, it may be
infinite in the large $N$ limit, and we can fix
${\alpha'}^2\sim g^2N^{1-\kappa}\ (\kappa>0)$ \cite{AIKKTT}.
In this case, however, we should impose the condition that at least
connected parts of the correlation functions of the (renormalized)
Wilson loops are finite in the large $N$ limit while various
correlation functions go to infinity.

Our analysis is applicable to the bosonic model (\ref{eqn:Bmodel})
since supersymmetry does not play an important role in counting
the powers of $N$ in our arguments.
One might think that the $N$ dependences of correlation functions
would be different in supersymmetric and bosonic models. However,
the distributions of the eigenvalues of the bosonic matrices in these
models agree with each other \cite{AIKKT,MCsimu}. This supports
our analysis at least in the leading order of $N$.
Other correlation functions, such as $\langle\frac{1}{g^2}\Tr
A^4\rangle$ and the Wilson loops, were previously calculated
in the bosonic matrix model using Monte Carlo simulations
\cite{MCsimu,alpha} and our results agree with those.
It is surprising that our two equations (\ref{eqn:scaling-b}) and
(\ref{eqn:scaling-f}) lead to the correct large $N$ behavior for
the correlation functions, such as $\langle\frac{1}{N}\Tr A^2\rangle$,
$\langle\frac{1}{g^2}\Tr A^4\rangle$ and the Wilson loops, when $g^2N$
is fixed. However, an analytical proof for validity of the
assumptions used to obtain the equations is not given here.
These results suggest that all the momentum modes of the fields, or the
components of the matrices, contribute significantly at least in the
bosonic model.

\begin{table}
\begin{center}
\begin{tabular}{l|l}
IIB matrix model ($S^M_{IIB}$) & Schild action ($S_1$)\\ \hline
$SU(N)$ & area preserving diffeomorphism  \\
$\begin{array}{l}
 \delta_\Lambda A^\mu=i[\Lambda,A^\mu]\\
 \delta_\Lambda \Psi~=i[\Lambda,\Psi]
 \end{array}$
	&$ \begin{array}{l}
	 \delta_{\lambda(\sigma)}A^\mu(\sigma)
	  =\epsilon^{ab}\,\partial_a\lambda(\sigma)\,\partial_b
	A^\mu(\sigma)\\
	 \delta_{\lambda(\sigma)}\Psi(\sigma)~
	  =\epsilon^{ab}\,\partial_a\lambda(\sigma)\,\partial_b
	\Psi(\sigma)
	 \end{array}$\\ \hline
supersymmetry & supersymmetry\\
$ \begin{array}{l}
 \delta_\varepsilon A^\mu=i {\overline\varepsilon}\Gamma^\mu\Psi\\
 \delta_\varepsilon \Psi~=
 \frac{i}{2}[A^\mu,A^\nu]\Gamma_{\mu\nu}\varepsilon
 \end{array}$
	&$ \begin{array}{l}
	 \delta_\varepsilon A^\mu(\sigma)
	  = i {\overline\varepsilon}\Gamma^\mu\Psi(\sigma)\\
	 \delta_\varepsilon \Psi(\sigma)
	  =\frac{i}{2}\epsilon^{ab}\partial_a
	  A^\mu(\sigma)\,\partial_b A^\nu(\sigma)\,
	   \Gamma_{\mu\nu}\varepsilon\label{eqn:SUSY}
	 \end{array}$ \\ \hline
rotational invariance & rotational invariance \\
$\begin{array}{l}
 \delta_\omega A^\mu=\omega^\mu_{~\nu}A^\nu\\
 \delta_\omega \Psi~=0\\
 (\omega_{\nu\mu}=-\omega_{\mu\nu})
 \end{array}$
	&$\begin{array}{l}
	 \delta_\omega A^\mu(\sigma)=\omega^\mu_{~\nu}A^\nu(\sigma)\\
	 \delta_\omega \Psi(\sigma)~=0\\
	 (\omega_{\nu\mu}=-\omega_{\mu\nu})
	\end{array}$\\ \hline
shift of the bosonic matrices & shift of the bosonic fields\\
$ \begin{array}{l}
 \delta_\alpha A^\mu=\alpha^\mu{\bf 1}\\
 \delta_\alpha \Psi~=0
 \end{array}$
	&$ \begin{array}{l}
	 \delta_\alpha A^\mu(\sigma)=\alpha^\mu\\
	 \delta_\alpha \Psi(\sigma)~=0
	 \end{array}$\\ \hline
shift of the fermionic matrices & shift of the fermionic fields\\
$ \begin{array}{l}
 \delta_\chi A^\mu=0\\
 \delta_\chi \Psi~=\chi{\bf 1}
 \end{array}$
	&$ \begin{array}{l}
	 \delta_\chi A^\mu(\sigma)=0\\
	 \delta_\chi \Psi(\sigma)~=\chi
	 \end{array}$ \\
\end{tabular}
\caption{Correspondence between symmetries of the IIB matrix model and
$S_1$}
\label{tab:symmetry}
\end{center}
\end{table}

If we regard the IIB matrix model for finite $N$ (\ref{eqn:SikktOrg})
as a regularized action of some field theory, the Schild action
\cite{Schild} is a plausible candidate for such a field theory.
Actually, the a naive large $N$ limit of the action of the field
theory on the NCPL (\ref{eqn:SikktNCT}) is the Schild action
(\ref{eqn:Sikkt}).
We have seen that the $N$ dependence of the action is crucial for
calculating the $N$ dependences of correlation functions.
The lattice-regularized Schild action (\ref{eqn:Sikkt-sum}), however,
has a different $N$ dependence from that of the IIB matrix model
(\ref{eqn:SikktNCT}). This difference comes from the fact that we
have neglected the $O(\frac{1}{N^2})$ terms in the commutation
relations in (\ref{eqn:lNcr}). This is justified only if large
momentum modes do not contribute to the correlation functions.
Thus the large momentum modes ${\bf m} = O(N)$
contribute equally or significantly to correlation functions,
since our results agree with those obtained by the matrix
model \cite{MCsimu,alpha}.

What sort of action can we have for the IIB matrix model in the
large $N$ limit?  Noting that $\partial_a = O(N)$ \cite{KU}, we
may consider the action
\begin{equation}
  S_1=\frac{1}{g^2N^3}\int_{T^2}\hspace{-1ex}d^2\sigma
	\left[\frac{1}{4}\left(\epsilon^{ab}\,
	\partial_aA^\mu(\sigma)\,\partial_bA^\nu(\sigma)\right)^2
	-\frac{i}{2}\,{\overline\Psi}(\sigma)\,\epsilon^{ab}\,
	\partial_a{\Slash A}(\sigma)\,\partial_b\Psi(\sigma)\right]\,.
	\label{eqn:Schild-2}
\end{equation}
This action is different from the Schild action (\ref{eqn:Sikkt}) by
the factor $N^{-2}$.
Although we can write down $S_1$ by rescaling the fields in the Schild
action (\ref{eqn:Sikkt}), we cannot obtain $S_1$ directly from the
action of the IIB matrix model (\ref{eqn:SikktOrg}).
$S_1$, however, is one of the most likely candidates for the
action of the IIB matrix model in the large $N$ limit because each
symmetry of the IIB matrix model corresponds to one of the symmetries
of $S_1$ (see Table \ref{tab:symmetry}).
On the other hand, we found that we cannot neglect
$O(\frac{1}{N^2})$ terms of the commutation relations (\ref{eqn:lNcr})
in the large $N$ limit. This implies that the large $N$ limit of a
field theory on an NCPL, or the matrix model, will be a theory on a
non-commutative torus or a non-local theory on a torus.
Non-commutative field theories have the special feature that there
exist stringy modes as well as usual particle modes \cite{NCG}.
This suggests that the large $N$ field theory on an NCPL
describes the Planck scale physics in which the coordinates
do not commute with each other, and hence it supports the
conjecture that the IIB matrix model describes the theory at the
Planck scale.

There may be other field theories which have the same $N$
dependence as the IIB matrix model, and one of them must be the
true action of the IIB matrix model in the large $N$ limit.
If we could find such an action, we would have much more information,
such as that regarding the true vacuum.
Searching for such an action is one of the most important subjects in
studying the IIB matrix model.

\section*{Acknowledgements}
The work of N.\ K.\ is supported in part by the Japan Society for the
Promotion of Science under the Predoctoral Research Program.

\end{document}